# Mathematical optimization of treatment schedules

Bram L. Gorissen PhD, Jan Unkelbach PhD and Thomas R. Bortfeld PhD[1]
Department of Radiation Oncology, Massachusetts General Hospital and Harvard Medical School.
[1] corresponding author. Massachusetts General Hospital, Department of Radiation Oncology, 100 Blossom Street Cox 362, Boston, MA 02114. Phone: 617-724-1180. E-mail: tbortfeld@mgh.harvard.edu

In the past decades mathematical optimization has found its way into radiation therapy and has made profound practice changing impact. Today, virtually all advanced treatment delivery methods, such as IMRT, VMAT, tomotherapy, LDR/HDR brachytherapy, proton therapy, are based on some form of optimization approach that changes the treatment variables (beam intensities, multileaf collimator shapes, beam angles, dwell times and seed or source positions) in the planning computer using an optimization algorithm, until the best value of the treatment (planning) objective has been found. It is fair to say though that a truly optimal radiation treatment plan remains elusive. The reasons for that include difficulties in defining meaningful planning objectives and constraints in mathematical terms, various uncertainties in the planning and delivery process, and the inability of optimization algorithms to find the true optimum. Radiation therapy optimization therefore remains an active field of research.

While a truly optimal treatment plan cannot generally be achieved, dose distributions today are significantly better than those we could deliver 25 years ago, i.e., before the widespread introduction of optimization techniques in our field (as well as many other recent advances). In fact, today our treatments are approaching the physical limits in terms of spatial dose conformation, certainly in the case of (external) photon therapy. With the ability to deliver much tighter dose distributions, it appears to be a good idea to revisit the dose fractionation problem, and to include the timing of the dose delivery in a systematic optimization approach (Kim et al. 2015, Unkelbach et al. 2013, Wein et al. 2000, Yang and Xing 2005).

Pioneering work in the field of dose schedule optimization for glioblastoma (GBM) was recently published in Cell (Leder et al. 2014, Michor and Beal 2015). The overall goal in (Leder et al. 2014) was to optimize the radiation treatment schedule, particularly for PDGF-driven GBM. A highly non-standard fractionation schedule, derived from a newly proposed biological model, showed impressively prolonged survival in mice. Despite this proven potential, the call for clinical trials (Michor and Beal 2015) is premature, due to the challenges associated with this type of approach.

The two main components of any approach for the optimization of the fractionation schedule are, first, a biological model that links the times and delivered doses of the individual dose fractions with a measure of the outcome, such as tumor size after the end of treatment. The second main component is the optimization algorithm that finds the treatment schedule that optimizes the outcome based on the underlying model.

The biological model of Leder et al. assumes two types of tumor cells: differentiated and stem-like cells. Stem-like cells are assumed to have a much slower growth rate. In addition, it is assumed that radiation triggers the conversion of fast growing differentiated cells to slowly dividing stem-like cells. Hence, it is desirable to schedule the radiation delivery such that it actually *maximizes* the number of stem-like cells (relative to the number of differentiated cells) in order to achieve growth delay.

The optimization approach of Leder et al. is based on simulated annealing, which tries a large number of randomly chosen schedules and picks the one with the smallest predicted tumor size at a fixed time after the end of treatment. Because the authors wanted to test their optimized schedule in mice, they limited the total dose to 10 Gy, to be delivered within one week, Monday to Friday, between 8am and 5pm in units of 1 Gy. Compared with the standard fractionation of 2 Gy per day, their optimized schedule demands different numbers of treatment fractions per day to be given at different times with different doses. Their highly irregular, non-standard optimized schedules led to significantly (53%) prolonged survival in the experiments with the mice bearing PDGF-driven GBM tumors.

Table 1: Overview of the schedules.

| Schedule | Day 1 | Day 2 | Day 3 | Day 4 | Day 5 |
|---|---|---|---|---|---|
| Optimum 1 (from Leder et al.) | 1 Gy 8am 2pm 5pm | 1 Gy 5pm | 1 Gy 3pm 5pm | 1 Gy 5pm | 1 Gy 3pm 4pm 5pm |
| Global Optimum 1 | 1 Gy 8am 5pm | 1 Gy 8am 5pm | 1 Gy 8am 5pm | 1 Gy 8am 5pm | 1 Gy 8am 5pm |
| Optimum 2 (from Leder et al.) | 3 Gy 8am | 1 Gy 4pm | | 1 Gy 9am 1pm 5pm | 1 Gy 9am 1pm 5pm |
| Global Optimum 2 | 2 Gy 8am 1 Gy 12pm 4pm | | 1 Gy 8am 12pm 4pm | | 1 Gy 8am 12pm 4pm |

Intrigued by those impressive results from Leder et al., we studied their biological model in order to understand the underlying assumptions that give rise to the non-intuitive fractionation schedules. In addition, we wanted to see if we can improve the results even further by using an exact optimization algorithm instead of the approximate simulated annealing method. In fact, it turns out that the mathematically optimal schedule can be found exactly (in three hours on a state-of-the-art desktop personal computer, single CPU core) through the method of exhaustive search (Appendix A). Table 1 compares the truly optimal schedule (*Global-Optimum-1*) with the schedule derived by Leder et al. (*Optimum-1*) for their original tumor model. Their schedule, which they tested experimentally, differs drastically from the schedule predicted to be best by the model, even though the survival would only be slightly improved (by 2%) with the globally optimal schedule. Leder et al. did notice inconsistencies with respect to predictions from their original tumor model. However, they did not question the optimization approach, but rather the model. As a consequence, they revised the model and added two parameters, which made the model agree better with existing data. Also for this revised model, we found the truly optimal treatment schedule (*Global-Optimum-2* in Table 1), and compared it with their schedule (*Optimum-2*). Again, this schedule is very different from the one found by Leder et al. As for *Global-Optimum-1*, survival is only slightly improved (by 10%) with the globally optimal schedule.

One of the most fascinating experimental results of the Leder et al. paper is that highly irregular fractionation schemes outperform standard schedules. However, as shown in Table 1, the fractionation schemes that the model truly predicts to be best are regular. The original model suggests a hyperfractionation scheme with two daily fractions nine hours apart (*Global-Optimum-1*). The revised model suggests a hyperfractionation scheme with three daily fractions four hours apart. Note that the minor irregularity (a 2 Gy fraction on Monday 8am) only arises because 10 fractions cannot be evenly divided over 3 days. Changing the model as to deliver 9 Gy in total instead of 10 Gy yields a perfectly regular schedule. Hence, the irregularity of the fractionation schemes found by the authors (*Optimum-1* and *2*) are only due to the use of simulated annealing, an approximate optimization method that artificially introduces randomness. The tumor model does not inherently suggest irregular fractionation schemes and cannot explain their superiority. This finding reminds us of the beginning of IMRT dose optimization, where simulated annealing was used for the optimization of the intensity maps, which then often looked "noisy" and were unnecessarily difficult to deliver using multileaf collimators.

Aside from sub-optimality due to insufficiently optimized dose schedules based on a given biological model, there are substantial uncertainties related to the biological model itself. Leder et al. took great care in deriving their model, iteratively improving the model and its parameters based on the results from the mouse experiments, and checking the stem cell predictions with side population (SP) analysis. However, some questions remain. The regular schedule that the revised Leder et al. model truly predicts to be optimal (*Global-Optimum 2*, three fractions per day, four hours apart), is solely determined by the assumption that the conversion from differentiated cells to slow growing stem-like cells is maximized when two fractions are delivered 3–4 hours apart (Appendix B). Realistically, there is very little evidence for the correctness of this assumption, aside from extrapolation of the model. Regular three-times-per-day schedules have been tested in the clinic with human GBM patients, and have not been found to yield better results than standard fractionation (Ludgate et al. 1988, Beşe et al. 1998, Lutterbach et al. 1999). However, repeating those previous studies specifically for patient populations with PDGF-driven GBM may still reveal improvements. This is clearly a testable hypothesis.

Overall, we believe that expanding the systematic optimization approach in radiation therapy, from the spatial optimization of dose distributions to the optimization of time dose fractionation, holds a lot of promise. However, it should be recognized that the role that mathematical optimization algorithms play in IMRT planning and fractionation schedule optimization is quite different. In IMRT planning, a physician determines beforehand, in terms of dose limits and prescription doses, how a patient is to be treated. The role of IMRT optimization is then to find a technical solution to the question how the desired dose distribution can be optimally approximated with external radiation beams. In contrast, optimizing fractionation schemes is hypothesis generating in nature. The goal is to propose novel treatment regimens that potentially improve outcome. Unlike spatial dose optimization, this requires a biological model to link the treatment schedule with outcome, the role of mathematical optimization is to provide the correct link between biological model assumptions and proposed treatment schedules. As in the case of IMRT optimization, one difficulty is that the optimization problem can become so big or so hard that a truly optimal solution may be impossible to find, and that approximate solutions may result in misleading artifacts, i.e., hypotheses that are not supported by the biological model. However, uncertainties related to the biological models pose another perhaps even more significant challenge. We agree with Michor and Beal (2015) that those challenges may indeed be surmountable – but much more effort will be necessary to make that happen.

## *Acknowledgments*

We thank Leo Gerweck and Helen Shih (MGH) for their thoughtful comments on a previous version of this manuscript.

# Online supplement

# A Determining the optimal fractionation scheme

In this section, we show how to determine the optimal fractionation schedule. The authors noted that "mathematically identifying the global optimal schedule was not computationally feasible due to the complexity of our model", and hence, resorted to simulated annealing to determine a fractionation scheme that significantly outperforms a standard scheme of 2 Gy per day in terms of life expectancy. Although the model is complex, we disagree that the globally optimal schedule cannot be found, and propose two alternative optimization methods that are guaranteed to find such a schedule.

**Exhaustive search** The number of ways in which 10 Gy can be delivered at 50 potential time moments with a maximum of three fractions per day is limited. If each Gy of dose is unique, there are $50^{10}$ ways of delivering 10 Gy in 50 time slots. After removing permutations, this reduces to $63 \cdot 10^9$. Additionally accounting for a maximum of three fractions per day leaves approximately $46 \cdot 10^9$ schedules, which a computer can enumerate in a moderate amount of time. The best regimen can be found by going over all possible fractionation schemes and computing the treatment outcome.

**Mixed integer programming** We define the state of a tumor prior to each fraction *i* as a triplet ($N_i^S$, $N_i^{dNR}$, $N_i^{dR}$), which are the number of SLRCs, the number of DSCs that are incapable of reversion, and the number of DSCs that may revert. This state definition satisfies the Markov property, which means that the future depends only on the current state and not on the past states. Knowing the state and the fraction dose $d_i$, the model has enough input information to describe the future dynamics of the tumor up to the next fraction. In particular, it holds that the number of SLRCs and DSCs prior to the next fraction are linear functions of $N_i^S, N_i^{dNR}$ and $N_i^{dR}$:

$$\begin{pmatrix} N_i^S \\ N_i^{dNR} \\ N_i^{dR} \end{pmatrix} = \begin{pmatrix} c_1 & c_2 & c_3 \\ c_4 & c_5 & c_6 \\ c_7 & c_8 & c_9 \end{pmatrix} \begin{pmatrix} N_i^S \\ N_i^{dNR} \\ N_i^{dR} \end{pmatrix},$$

where the constants $c_i$ depend on $d_i$ and the time between fractions *i* and *i+1*. Since $d_i$ is integral and between 1 and 10, it can take only 10 distinct values. Similarly, the time between fractions can only take a finite number of values. All possible values for the constants $c_i$ can therefore be precomputed, after which most of the complexity of the model disappears. Mixed integer optimization then offers the flexibility to describe the tumor cell population in an optimization framework, after which state-of-the-art solvers can be used to determine the optimal fractionation scheme. This has the potential to scale better than the exhaustive search approach when more dose is delivered over a longer time frame.

**Our implementation** The paper contains all information necessary to implement the model, with two exceptions. The first is the time moment at which a fractionation scheme was assessed. The composition of the tumor (the SLRC/DSC ratio) influences the tumor growth, so some effects may only appear in the long run. The authors have communicated that the schedules were compared 14 days after treatment for determining *Optimum-1*, while for *Optimum-2* a time horizon of 5 or 6 weeks was used. The second piece of unclear information is what happens prior to the first fraction. We assume that the cells are not quiescent, because they have not been irradiated yet. Consequently, we assume that cells are proliferating up to the first fraction. We also assume that DSCs that may revert have already done so.

We have implemented exhaustive search, which can evaluate all possible fractionation schemes in approximately three hours. The corresponding source code accompanies this manuscript. We name the best fractionation schemes *Global-Optimum-1* and *2* for the initial and refined model, respectively (see Table 1). *Global-Optimum-2* is not unique, e.g., the schedule on day 3 may be shifted one hour ahead, or moved to day 2 or 4 to get an equivalent outcome. The schedule on day 5 may be moved to day 2 if the evaluation moment of "*n* days after treatment" is interpreted as *n* days after the last fraction (compared to *n* days after Friday 5pm). These globally optimal solutions allow us to assess the suboptimality of *Optimum-1* and *2*, and to further analyze the model.

# B Analysis of the model

In the model, the tumor is assumed to consist of two types of cells: stem-like resistant cells (SLRCs) and differentiated sensitive cells (DSCs). DSCs are more susceptible to radiation than SLRCs (higher $\alpha$ and $\beta$ but same $\alpha/\beta$ ratio). After each fraction, all cells become quiescent. Then, after a fixed amount of time (which for SLRCs is longer than for DSCs), the cells gradually become active and start proliferating at a given rate (which for DSCs is at least fifty times higher than for SLRCs). SLRCs can convert to DSCs at a certain rate even if they are quiescent, and vice versa, a fraction of the DSCs can revert to SLRCs.

The main hypothesis behind the model is that radiation triggers the conversion of DSCs to SLRCs. Since SLRCs are assumed to proliferate much slower than DSCs, growth delay (or improvements in median survival) can be achieved through fractionation schemes that convert as many DSCs to SLRCs as possible. The globally optimal schedules can be explained in that context together with the following considerations.

The globally optimal solutions mostly give fractions of 1 Gy each. Cell kill in the model is described by the LQ-model, where the $\alpha/\beta$ ratio is of the order of $10^6$. The quadratic effect is therefore negligible, so higher fraction doses do not pay off.

The model parameters that are used, are such that most effects that have been modeled only take place after the final fraction. As long as one fraction is given every 24 hours (refined model: 313 hours), the cells are quiescent between fractions, i.e., they only proliferate after the treatment. Therefore, the tumor dynamics during the course of treatment other than cell kill are exclusively described by the conversion from SLRCs to DSCs and the reverse process. The conversion rate from SLRCs to DSCs is so low that less than 2% of the SLRCs convert in a week, while 68% of the DSCs revert to SLRCs each hour.

For the initial model, 15% of the DSCs can revert, and those get reselected after each fraction. Suppose there are 1000 DSCs, then three hours later 145 out of the DSCs that can revert have done so (based on the assumed exponential conversion over time). When a new fraction is delivered after those three hours, again 15% out of the remaining 1000-145 = 855 can revert and almost all have done so three hours later. Without the new fraction, there would never have been less than 850 DSCs. Since each fraction triggers the reversion process independent of the dose level, it is beneficial to have 10 fractions of 1 Gy each, spaced maximally apart.

For the refined model, the same analysis holds, except that the fraction that can revert depends on the time between the previous two fractions. When they are more than seven hours apart, almost no DSCs are capable of reverting. Reversion is maximal for a 3 or 4 hour time gap, which suggests three daily 1 Gy fractions 3–4 hours apart. Since there are ten fractions to be divided over five days, there is one leftover fraction. For *Global–Optimum-2*, the effect of a single fraction on day 4 or 5 causes virtually no additional reversion ($\leq 10^{-24}$%), which explains why the quadratic effect still dictates a 2 Gy fraction at 8am on Day 1.

In summary, since DSCs are assumed to proliferate much quicker than SLRCs after exiting quiescence, the optimal fractionation schedule should primarily boost the reversion process and minimize the number of fast proliferating DSCs at the end of treatment. To this end, the features of the optimal solution are:
1. Having a large number of fractions that deliver the smallest dose allowed (1 Gy). This results from the assumption that classical fractionation effects are absent ($\beta \approx 0$), and that each fraction triggers the reversion process independent of the dose level.
2. For the initial model, it is optimal to spead out the 10 fractions uniformly as much as possible, resulting in a hyperfractionated schedule with bi-daily fractions that are delivered in the early morning and late afternoon. This is directly apparent from the model equations, and is confirmed by exhaustive search.
3. For the refined model, it is optimal to deliver 1 Gy fractions at time intervals of 3–4 hours. This directly follows from the assumption that cell reversion peaks after 3–4 hours, and a change in this model parameter would change the optimal fractionation scheme accordingly. Together with the working hour constraints (8am–5pm) and the limitation of at most 3 fractions per day, this results in 3 daily 1 Gy fractions 4 hours apart.